\documentclass{emulateapj}
\usepackage{verbatim} 

\shorttitle{Radio-FIR correlation}
\shortauthors{I.~Mori\'{c} et al.}

\def\f#1   {Fig.~\ref{#1}}
\def\s#1   {Sec.~\ref{#1}}
\def\tab#1   {Tab.~\ref{#1}}
\def\t#1   {Tab.~\ref{#1}}
\def\lum   {$\mathrm{L}_\mathrm{1.4GHz}$}
\def\comm#1   {{\tt (COMMENT: #1) }}

\def\wh                {W~Hz$^{-1}$}

\def\smo               {Smol\v{c}i\'{c}}

\slugcomment{  }

\begin{document}

\title{ A closer view of the radio-FIR correlation: Disentangling the contributions of star formation and AGN activity\altaffilmark{0}}

\author{I.~Mori\'{c}\altaffilmark{1,2}, 
        V.~Smol\v{c}i\'{c}\altaffilmark{3,4,2}, 
	A.~Kimball\altaffilmark{5,6},
	D.~A.~Riechers\altaffilmark{2,7},
	\v{Z}~Ivezi\'{c}\altaffilmark{5},
	N.~Scoville\altaffilmark{2}
        }
\altaffiltext{0}{Based on observations with the National
Radio Astronomy Observatory which is a facility of the National Science
Foundation operated under cooperative agreement by Associated Universities,
Inc. }
\altaffiltext{1}{University of Zagreb, Physics Department,
  Bijeni\v{c}ka cesta 32, 10002  Zagreb, Croatia}
\altaffiltext{2}{ California Institute of Technology, MC 249-17, 1200 East
California Boulevard, Pasadena, CA 91125 }
\altaffiltext{3}{ESO ALMA COFUND Fellow, European Southern Observatory, Karl-Schwarzschild-Strasse 2, 
85748 Garching b. Muenchen, Germany}
\altaffiltext{4}{Argelander Institut for Astronomy, Auf dem H\"{u}gel 71, Bonn, 53121, Germany}
\altaffiltext{5}{Department of Astronomy, University of Washington, Box 351580,
  Seattle, WA 98195-1580}
\altaffiltext{6}{National Radio Astronomy Observatory, 520 Edgemont Road,
Charlottesville, VA 22903}
\altaffiltext{7}{Hubble Fellow }

\begin{abstract}
  We extend the Unified Radio Catalog, a catalog of sources detected
  by various (NVSS, FIRST, WENSS, GB6) radio surveys, and SDSS, to IR
  wavelengths by matching it to the IRAS Point and Faint Source
  catalogs. By fitting each NVSS-selected galaxy's NUV-NIR spectral
  energy distribution (SED) with stellar population synthesis models
  we add to the catalog star formation rates, stellar masses, and
  attenuations. We further add information about optical emission line
  properties for NVSS-selected galaxies with available SDSS
  spectroscopy. Using an NVSS 20~cm
  ($F_\mathrm{1.4GHz}\gtrsim2.5$~mJy) selected sample, matched to the
  SDSS spectroscopic (``main'' galaxy and quasar) catalogs and IRAS
  data ($0.04<z\lesssim0.2$) we perform an in depth analysis of the
  radio-FIR correlation for various types of galaxies, separated into
  i) quasars, ii) star forming, iii) composite, iv) Seyfert, v) LINER
  and vi) absorption line galaxies using the standard optical
  spectroscopic diagnostic tools. We utilize SED-based star formation
  rates to independently quantify the source of radio and FIR emission
  in our galaxies.  Our results show that Seyfert galaxies have
  FIR/radio ratios lower than, but still within the scatter of, the
  canonical value due to an additional (likely AGN) contribution to
  their radio continuum emission. Furthermore, IR-detected absorption
  and LINER galaxies are on average strongly dominated by AGN activity
  in both their FIR and radio emission; however their average
  FIR/radio ratio is consistent with that expected for star forming
  galaxies.  In summary, we find that most AGN-containing
    galaxies in our NVSS-IRAS-SDSS sample have FIR/radio flux ratios
    indistinguishable from those of the star-forming galaxies that
    define the radio-FIR correlation. Thus, attempts to separate AGNs
    from star-forming galaxies by their FIR/radio flux ratios alone
    can separate only a small fraction of the AGNs, such as the
    radio-loud quasars.



\end{abstract}


\section{Introduction}
\label{sec:introduction}

The radio-FIR correlation is one of the tightest correlations in
observational astrophysics (e.g.\ \citealt{helou85, condon92, mauch07,
  yun01, bell03, sargent10, kovacs06, murphy09, appleton04}). The
correspondence between the radiation in the (far-)infrared and that in
the radio spans over nearly five orders of magnitude in various types
of galaxies, ranging from dwarfs to ULIRGs.  Given that the two
observational windows, IR and radio, trace independent and different
intrinsic physical mechanisms in galaxies -- thermal vs.\ synchrotron
radiation -- the existence of such a tight correspondence is
remarkable. It is generally believed that recent star formation in
galaxies is the process that relates IR and radio emission.  

The radio-FIR correlation has been extensively studied in the past
both in the low \citep{helou85, condon92, garrett02, mauch07, yun01,
  bell03} and high redshift universe \citep{sargent10, kovacs06,
  sajina08, murphy09, appleton04, vlahakis08, ibar08, chapman05}.
It has been shown that, out to redshifts of $z\sim3-4$, the
  FIR/radio ratios of various types of galaxies are essentially the
  same as those in the local universe.  At higher redshifts radio
  quiet QSOs have been demonstrated to have FIR/radio ratios
  consistent with the local value, while the FIR/radio ratios of $z>4$
  SMGs are found to be lower by a few factors.  This is somewhat
  contrary to expectations, as the the FIR/radio ratio is expected to
  be rising with redshift (especially at $z\gtrsim3$) due to the
  increase of the cosmic microwave background (CMB) energy density
  ($U_\mathrm{CMB}$) with redshift, $U_\mathrm{CMB}\propto(1+z)^4$,
  which surpresses the non-thermal component of a galaxy's radio
  continuum via inverse-Compton (IC) scattering (see Murphy 2009 for
  details).  An explanation for this discrepancy can be provided by
  additional processes that add to a galaxy's radio continuum, such as
  increased magnetic field strengths or AGN contribution, that may
  compensate for the radio continuum emission losses due to IC
  scattering.  

The AGN contribution to the radio-FIR correlation has been studied in
the past to some extent. Typically a low FIR/radio ratio,
significantly offsetting a galaxy from the correlation, is thought to
indicate a radio-loud AGN \citep[e.g.][]{yun01,condon02}. However,
recent studies have shown that optically-selected AGN often follow the
correlation, albeit with a slightly lower FIR/radio ratio. For
example, based on SDSS-NVSS-IRAS data, \citet{obric06} have
demonstrated a tight correlation between radio and 60~$\mu$m fluxes
for low-luminosity AGN (predominantly Seyferts and LINERs), which
varies by only $\sim20\%$ relative to that of star forming
galaxies. Utilizing 6dFGS-NVSS-IRAS data, \citet{mauch07} inferred a
lower average FIR/radio ratio for AGN-bearing galaxies (Seyferts,
LINERs, and quasars), but still within the scatter of the correlation
for star forming galaxies.  Furthermore, studies of the correlation at
higher redshifts have yielded a handful of interesting objects for
which it has clearly been shown that a significant AGN contribution to
IR and/or radio exists, yet their FIR/radio ratio is consistent with
the canonical value for star forming galaxies (Riechers et al.\ 2009,
Murphy et al.\ 2009).

In order to understand in more detail the contribution of AGN activity
to the radio-FIR correlation, we perform an in-depth study of the
radio-FIR correlation, with a large sample, as a function of galaxy
type, and comparison with star formation rates for those individual
samples.  The various types of star forming and AGN-bearing galaxies
have been drawn from the NVSS \citep{condon98}, IRAS
\citep{neugebauer84}, and SDSS \citep{york00} sky surveys. In
\s{sec:data} \ we present the data used in this paper. We present the
correlation for various types of galaxies in \s{sec:quant} . In
\s{sec:lexcess} \ we link the FIR and radio emission from galaxies in
our sample to independently derived star formation rates, and in
\s{sec:discussion} \ and \s{sec:summary} \ we discuss and summarize
our results, respectively. We adopt $H_0=70$, $\Omega_M=0.3$,
$\Omega_\Lambda=0.7$, and define the radio synchrotron spectrum as
$F_\nu\propto\nu^{-\alpha}$, assuming $\alpha=0.7$. Throughout
  the text we will often use the term 'quasar' referring to both
  quasi-stellar radio sources and quasi-stellar objects.

\section{Data and galaxy samples: Expanding the Unified Radio Catalog}

\label{sec:data}


\subsection{Unified Radio Catalog}

\citet{kim08} have constructed a catalog of radio sources detected by
the GB6 (6~cm), FIRST \citep{becker95}, NVSS (\citealt{condon98};
20~cm), and WENSS (92~cm) radio surveys, as well as the SDSS (DR6)
optical survey \citep{york00}. This ``Unified Radio Catalog'' has been
generated in such a way that it allows a broad range of 20~cm based
sample selections and source analysis (see \citealt{kim08} for
details).  The 2.7 million entries are comprised of the closest three
FIRST to NVSS matches (within $30''$) and vice-versa, as well as
unmatched sources from each survey. All entries have been supplemented
by data from the other radio and optical surveys, where available.
Here we select from the Unified Radio Catalog (version 1.1) all 20~cm
sources that have been detected by the NVSS radio survey (using
$\mathrm{matchflag\_nvss}=-1$ and $\mathrm{matchflag\_first}\leq1$;
see \citealt{kim08} for details). This selection yields a radio flux
limited ($F_\mathrm{1.4GHz}\gtrsim2.5$~mJy) sample that contains
1,814,748 galaxies. In the following section we expand this catalog to
IR wavelengths, and augment it with additional (spectroscopic and
SED-based) information.

\subsection{Expanding the Unified Radio Catalog}

\subsubsection{IRAS}

For the purpose of this paper, we have expanded the Unified Radio
Catalog to IR wavelengths by cross-correlating it with the IRAS
point-source and faint-source catalogs (hereafter PSC and FSC, resp.).
The IRAS PSC contains 245,889 confirmed point sources detected at 12,
25, 60 and 100 $\mu$m, respectively \citet{strauss90}. The
completeness of the catalog at these wavelengths reaches down to 0.4,
0.5, 0.6 and 1.0 Jy, respectively. The FSC was tuned to fainter levels
based on the same IR data by point-source filtering the individual
detector data streams and then coadding those using a trimmed-average
algorithm (see \citealt{moshir92}).  The reliability of the FSC is
slightly lower than that of the PSC ($\gtrsim94\%$ compared to
99.997$\%$); however its sensitivity is higher by a factor of
$\sim2.5$. The FSC contains 173,044 point sources with flux densities
typically greater than 0.2 Jy at 12, 25 and 60~$\mu$m and 0.5 Jy at
100~$\mu$m.



We used matching radius of $30"$, as optimized by \citet{obric06}, in cross-corelating the Unified Radio Catalog with the IR IRAS data.
In \f{fig:Distance_IRAS_NVSS} \ we show the distribution of
the distances between the IR and radio detections. The cumulative
distribution displayed in \f{fig:Distance_IRAS_NVSS} \ shows that
$\sim70\%$ of the positional matches are within an angular distance of
$15''$.

Our NVSS-selected radio sample contains 18,313 galaxies with high
quality IR photometry\footnote{We take the IRAS quality indicator,
  reported in the FSC and PSC, to be $\geq2$ at 60 and 100~$\mu$m
  (the wavelength bands utilized here).  } (see \t{table:class} ). As
the FSC and PSC have been generated based on the same data, most of
the PSC sources are included in the FSC. In our entire NVSS-IRAS
sample, 26\% of the sources have a PSC detection but are not included
in the FSC. This fraction, however, reduces to only 3\% after an
optical (SDSS) cross-match is performed.
%
%

The 60 and 100~$\mu$m magnitudes reported in the PSC and FSC are in
agreement for the union of the two IR samples. The biweighted mean of
the flux difference (for a subsample with SDSS detections) is 0.02 and
0.03~Jy at 60 and 100~$\micron$, respectively. The root mean scatter
of the 60~$\micron$ flux difference distribution is 0.06~Jy, while
that of the 100~$\micron$ distribution is significantly larger, i.e.\
0.16~Jy. Therefore, in order to access the highest quality IR
photometry, hereafter we use the values reported in either the FSC or
PSC catalog corresponding to the higher photometric quality flag
quoted in the catalogs.  The distribution of the 60~$\mu$m and 20~cm
flux densities are shown in \f{fig:LogFnvss} .

\subsubsection{SDSS quasar and main galaxy sample catalogs}

We have further matched the NVSS-selected sample from the Unified
Radio Catalog with data drawn from i) the SDSS DR5 quasar sample
\citep{schneider07}, and ii) the DR4 ``main'' spectroscopic sample for
which derivations of emission line fluxes from the SDSS spectra are
available (see \smo\ 2009 and references therein; note that the DR5
quasar and DR4 main galaxy catalogs were the most up-to-date versions
available at the time). The latter was complemented with stellar
masses, star formation rates, dust attenuations, ages, metallicities,
and a variety of other parameters based on spectral energy
distribution (SED) fitting of the SDSS ($ugriz$) photometry using the
\citet{bruzual03} stellar population synthesis models. The SED fitting
was performed as described in detail in \smo\ et al.\ (2008).

During the inspection of the validity of the final catalog, we have
found about 1$\%$ of objects with different spectroscopic redshifts in
various SDSS data releases ($\Delta z > 5\times10^{-4}$). We have
excluded those from the sample. Furthermore, a small number ($\sim
0.2\%$) of duplicate objects was present in both the SDSS ``main''
galaxy sample and the SDSS Quasar Catalog. Visually inspecting their
spectra yielded that most of these objects are better matched to the
properties of the ``main'' galaxy sample (as no power-law continuum
nor broad emission-lines were present in the spectrum), and we have
excluded these from our quasar sample. A summary of the various
radio-IR-optical samples is given in \t{table:class} , and in
\f{fig:limits1} \ and \f{fig:limits2} \ we show the radio (20~cm),
optical ($r$ band), and far-IR luminosities as a function of redshift
for the final NVSS-SDSS and NVSS-SDSS-IRAS samples (see
eqs.~\ref{for:lfir} and \ref{for:lr}). Note that the shallow IRAS
sensitivity (compared to the NVSS and SDSS data) significantly reduces
the number of objects, and biases the sample toward lower redshifts.

\subsection{Radio-optical-IR samples}

\subsubsection{Star forming and AGN galaxy subsamples}
\label{sec:bptselection}

We have used the optical spectroscopic information added to the NVSS
selected sample to spectroscopically separate the galaxies present in
the SDSS (DR4) ``main'' galaxy sample as absorption line, AGN
(LINER/Seyfert), star forming, or composite galaxies.

We define emission-line galaxies as those where the relevant emission
lines (H$\alpha$, H$\beta$, O[III,$\lambda$5007], N[II,$\lambda$6584],
S[II,$\lambda$$\lambda$6717,6731]) have been detected at
$\mathrm{S/N}\geq3$, and consider all galaxies with $\mathrm{S/N}<3$
in any of these lines as absorption line systems (see e.g.\
\citealt{best05, kewley06, smo09}).  As strong emission
lines are not present in the spectra of the latter, yet they are
luminous at 20~cm, they can be considered to be (low excitation) AGN
(see e.g.\ \citealt{best05,smo08} for a more detailed discussion).
Further, as illustrated in \f{fig:bpts} , using standard optical
spectroscopic diagnostics \citep{bpt81,kauffmann03b, kewley01,
  kewley06} we sort the emission-line galaxies into 1) star forming,
2) composite, 3) Seyfert, and 4) LINER galaxies.
The last two classes have been selected `unambiguously' by requiring
combined criteria using three emission line flux ratios (see middle and
right panels in \f{fig:bpts} ). A summary of the number of objects in
each class is given in \t{table:class} . It is noteworthy that the IR
detection fraction is a strong function of spectral class. It is the
lowest for absorption line (0.6\%) and LINER (6.5\%) galaxies,
intermediate for Seyferts (22\%) and the highest for composite (40\%)
and star forming (46\%) galaxies. These results suggest lower amounts of dust
(and gas; \citealt{solomon95}) in the former or alternatively
dominantly very cold dust that peaks at longer wavelenghts. 
  
The redshift distribution of the various galaxy types with 20~cm NVSS
and NVSS-IRAS detections is shown in the two top panels in
\f{fig:zhist} . Note that the redshift distribution of 20~cm detected
absorption line galaxies is biased toward higher-redshifts, compared
to all other galaxy types (see top panel in \f{fig:zhist} ). However,
this is not the case when an IRAS IR detection is required, as
illustrated in the middle panel in \f{fig:zhist} . The IR detection
fraction of the different galaxy classes is shown as a function of
redshift in the bottom panel in \f{fig:zhist} . Except for the overall
trend that absorption and LINER galaxies are detected less efficiently
in the IR, there is no substantial difference between the detection
fractions as a function of redshift for different types of
spectroscopically selected galaxies.

Hereafter, we apply redshift range limits of 0.04 $<$ z $<$ 0.3 to our
sample. The lower redshift limit is adopted from
\citet{kewley05}. Kewley et al.\ explored effects of fixed-size
aperture of the SDSS spectroscopic fibers on the spectral
characteristics such as metallicity, star formation rate, and
reddening. They concluded that a minimum aperture size covering
$\approx20\%$ of spectral light was required to properly approximate
global values. The SDSS fiber aperture of $3''$ diameter collects such
a fraction of light for galaxies of average size, type, and
luminosity at $z\gtrsim0.04$.  The upper redshift limit of $z=0.3$ is
equivalent to that of the SDSS ``main'' spectroscopic sample (note
however that the majority of IR-detected galaxies are at $z<0.2$, see
\f{fig:limits2} ).  It is worth noting that, because of lower spectral
signal-to-noise for fixed-luminosity galaxies at greater distances,
galaxies with weak emission lines, such as LINERs, can get confused with
absorption line galaxies at $z>0.1$ \citep{kewley05}. However, as
LINER and absorption galaxies have similar physical properties,
(e.g.\ \smo\ 2009) we simply combine these two types of galaxies,
and treat them hereafter as a single class.

\subsubsection{Quasar subsample}

Matching the SDSS DR5 quasar catalog to the Unified radio catalog
resulted in 4,490 matches (see \t{table:class} ).  The redshift range of
our radio luminous quasars is $0.09-5.12$, with a median at
$z=1.36$. Requiring IRAS detections biases the sample toward low
redshifts ($0.12\leq z\leq1.15$), with a median redshift of $0.18$,
and selects only $\sim0.5\%$ of the radio detected quasars.
The radio ($\gtrsim10^{23}$~\wh ) and FIR
($\gtrsim2\times10^{11}~\mathrm{L_\odot}$) luminosities (see
eqs.~\ref{for:lfir} and \ref{for:lr}) of our quasars are
systematically higher than those of the SDSS ``main'' spectroscopic
sample galaxies in our radio-optical-IR sample.

\section{Quantifying the radio-FIR correlation for various source types}
\label{sec:quant}

\subsection{Parameterizing the radio-FIR correlation}

The radio-FIR correlation is usually quantified by its slope via the
$\textit{q}$ parameter \citep{helou85}, defined as the logarithmic
ratio of the far-infrared flux to radio flux density:

\begin{equation}
q = \log \left\{\frac{F_\mathrm{FIR}/(3.75 \times 10^{12} \,\mathrm{Hz})}{F_\mathrm{1.4GHz}}\right\}
\label{for:q}
\end{equation}

where $F_\mathrm{1.4GHz}$ is the 1.4 GHz radio flux density in units of $\mathrm{W m^{-2} Hz^{-1}}$ and
$F_\mathrm{FIR}$ is the far-infrared flux in units of $\mathrm{W m^{-2}}$. Following \citet{sanders96}, we define the latter as: 

\begin{equation}
F_\mathrm{FIR} = 1.26 \times 10^{-14}\, ( 2.58 S_{60\micron} + S_{100\micron}) 
\label{for:fir}
\end{equation} 

where $S_{60\micron}$ and $S_{100\micron}$  are observed flux
densities at 60 and 100$\micron$ (in Jy), respectively.


We compute the far-infrared luminosity as:

\begin{equation}
L_{FIR} = 4\pi D_{L}^{2} C F_{FIR}\left[L\odot\right]
\label{for:lfir}
\end{equation}

where $D_{L}$ is the luminosity distance (in units of m) and
$C$ is a scale factor used to correct for the extrapolated flux
longward of the IRAS 100~$\mu$m filter. We use $C=1.6$ (see Tab.~1 in
\citealt{sanders96}).  Note that this expression can also be utilized
to compute the FIR luminosities for our IR-detected quasars, given
their relatively low redshifts.

The radio luminosity density is computed as:

\begin{equation}
L_\mathrm{1.4GHz} = \frac{4\pi D_{L}^{2}}{(1+z)^{1-\alpha}}F_\mathrm{1.4GHz}
\label{for:lr}
\end{equation}

where $z$ is the redshift of the source, $F_\mathrm{1.4GHz}$ is its
integrated flux density, and $\alpha$ the radio spectral index
(assuming $F_\nu\propto\nu^{-\alpha}$). To compute the radio luminosities, we
assumed a spectral index of $\alpha=0.7$.


\subsection{Radio-FIR correlation for all sources}

The radio-FIR correlation for the NVSS-SDSS-IRAS sample is summarized
in \f{fig:IRradio} . The radio and FIR flux densities (top left panel)
and luminosities (top right panel) clearly show a tight correlation
that holds over many orders of magnitude.  In the middle panels we
show the $q$ parameter, that characterizes the slope of the radio-FIR
correlation (see eq.~\ref{for:q}), as a function of FIR and radio
luminosities.  The average $q$ is constant as a function of FIR
luminosity (middle left panel), and it is decreasing with increasing
radio power (middle right panel; see also below).  In the bottom
panels of \f{fig:IRradio} \ we show the $q$ parameter as a function of
redshift, as well as its distribution for all our NVSS-SDSS-IRAS
sources (galaxies and quasars). We find that the average (biweighted
mean) $q$ value for the entire NVSS-SDSS-IRAS sample is
q=2.273$\pm$0.008, with a root-mean-square scatter of
$\sigma=0.18$. This is in very good agreement with previous findings
\citep{condon92,yun01,condon02,bell03,mauch07}, and will be discussed
in more detail in \s{sec:discussion} .

The quasars in our sample comprise the high luminosity end at both IR
and radio wavelengths (they are also located at higher redshifts,
compared to the IR and radio detected ``main'' galaxy sample). It is
also obvious that there is a larger fraction of quasars that do not
lie on the radio-IR correlation, compared to that for the ``main''
sample galaxies. 

\subsection{Radio-FIR correlation for different types of galaxies}
\label{sec:gals}

In \f{fig:FIRradio2} \ we present the radio-FIR correlation for the
SDSS ``main'' galaxy sample subdivided into different,
spectroscopically selected galaxy types (absorption, LINER, Seyfert,
composite and star forming galaxies; see \s{sec:bptselection} \ and
\f{fig:bpts} \ for details on the selection).  The decrease in
  $q$ with increasing radio power for all types of galaxies (middle
  right panel) is consistent with various other observations (e.g.\
  \smo\ et al.\ 2008; Kartaltepe et al.\ 2010; Sargent et al.\ 2010,
  Ivison et al.\ 2010). Note that, given the definition of the
  FIR/radio ratio, for any sample in which $q$ does not vary with FIR
  luminosity, and has a non-zero dispersion, it is expected to
  decrease with increasing radio luminosity (see e.g.\ Condon
  1984). To test whether the magnitude of the decrease is as expected
  from statistics or e.g.\ higher due to an additional effect (such as
  AGN contribution) we have computed the radio luminosity for each
  source based on its observed FIR luminosity and a FIR/radio ratio
  drawn from a Gaussian distribution with a dispersion of 0.18 and a
  mean of 2.27. We find that the decrease of $q$ with radio luminosity
  in the observed data is consitent with that in the simulated data,
  thus not requiring additional effects (such as increasing AGN
  contribution with increasing radio power) to explain this trend (at
  least in the radio luminosity range probed here).

A quantitative analysis of the radio-FIR
correlation for different galaxy types is presented in Fig
$\ref{fig:param}$. The spectroscopic selection of pure star forming
galaxies allows us to quantify the radio-IR correlation in a rather
unbiased manner. For our star forming galaxies we find an average $q$
value of $<q>=2.27\pm0.05$, with a small root-mean-square scatter of
$\sigma=0.13$. It is interesting to note that as the AGN contribution
rises in galaxies (as inferred based on optical spectroscopic
diagnostics) the scatter in $q$ increases by $\sim50\%$ to
$\sim150\%$. Interestingly, the scatter is the highest for Seyfert
types of galaxies, for which we also find the lowest average $q$
value, $<q>=2.14\pm0.05$. These differences will further be discussed
in \s{sec:discussion} .

\subsection{Radio-FIR correlation for quasars}
\label{sec:qso}

In \f{fig:FIRradioQSO} \ we quantify the radio-FIR correlation for the
21 IR-detected quasars in our sample. The distribution of the
FIR/radio ratio cannot be well fit with a Gaussian distribution. The
median $q$ value of the sample is 2.04, comparable to the average $q$
value we have found for Seyfert galaxies (2.14), and lower than that
for star forming galaxies (2.27; see \f{fig:param} ). It is worth
noting that the higher redshift quasars ($0.2\lesssim z\lesssim0.4$)
appear to be biased toward more radio-loud AGN.

\section{An independent view of the radio-FIR correlation: A link to star formation}
\label{sec:lexcess}

It is generally taken that recent star formation drives both the radio
and FIR emission of galaxies that lie on the radio-FIR correlation
\citep{condon92,mauch07}. Therefore, a correlation is expected to be
present between the SFRs and radio/FIR luminosities obtained from the
fluxes of galaxies dominated by recent star formation. To shed light
on the source of radio/FIR emission in our galaxies, in this section
we investigate the correlation between their radio/FIR luminosities
and star formation rates, independently determined based on fitting
stellar population synthesis models to the NUV-NIR SED).

We have derived a star formation rate for every galaxy in our sample by
fitting the \citet{bruzual03} library of stellar population synthesis
models to the SDSS $ugriz$ photometry (see \s{sec:data} ).  In
\f{fig:SFRmain} \ we show the radio and FIR luminosities of our sources
as a function of our SED-based SFRs. As expected, a correlation is
discernible between these two quantities. This is especially
emphasized for star formation dominated galaxies (i.e.\ star forming
and composite galaxies), the distribution of which agrees well with
the commonly used radio/IR luminosity -- SFR calibrations
\citep{kennicutt98, yun01}. Note that this is quite remarkable as the
SFRs have been derived completely independently from the FIR or radio
emissions in the galaxies.

From \f{fig:SFRmain} \ it is obvious that a large fraction of galaxies
with significant AGN contribution (Seyfert, LINER and absorption
galaxies) has an obvious excess of radio power and FIR luminosity
compared to that expected from the galaxy's SFR. The most obvious
example of this are the LINER and absorption galaxies from both the
NVSS-SDSS and NVSS-SDSS-IRAS samples.

To investigate whether star formation is the underlying source of
radio/FIR emission in our galaxies, we further quantify the difference
between the radio/FIR emission and that expected from star
formation. We thus define an ``excess'' in radio/FIR emission relative
to that expected from star formation ($\Delta \log{L_\mathrm{1.4Ghz}}$
and $\Delta \log{L_\mathrm{FIR}}$, resp.)  as:

\begin{equation}
\Delta \log{L}= \log{L_\mathrm{data}} - \log{L_\mathrm{exp.}}
\label{for:delta}
\end{equation}

where $\log{L_\mathrm{data}}$ is the logarithm of the 1.4~GHz or FIR
luminosity derived based on NVSS or IRAS data (see \s{sec:data} ), and
$\log{L_\mathrm{exp.}}$ is the luminosity (either at 1.4~GHz or FIR)
expected based on the SED-derived SFR and the standard radio and FIR
luminosity to SFR calibrations. To convert SFR to radio luminosity we
use the calibration defined in \citet{yun01}:
$\mathrm{SFR\,[M_\odot/yr]}=5.9\times10^{-22}\,L_\mathrm{1.4GHz}$~[W/Hz]. To convert SFR to FIR luminosity we use the
standard conversion defined by Kennicutt (1998):
$\mathrm{SFR\,[M_\odot/yr]}=4.5\times10^{-37}\,L_\mathrm{FIR}$~[W].
Prior to applying these conversions, derived using a Salpeter IMF, we
have scaled our SED-based SFRs by -0.2~dex to convert from a Chabrier
to a Salpeter IMF (we have additionally included a scaling factor of
$\sim0.4$~dex to account for the star formation histories used in our
models; see \smo\ et al.\ 2008 and Walcher et al.\ 2008 for details).

\f{fig:SFRmainA} \ shows the FIR ($\Delta \log\mathrm{L_{FIR}}$)
versus 1.4~GHz ($\Delta \log\mathrm{L_{1.4GHz}}$) luminosity excess
for different types of galaxies in the NVSS-SDSS-IRAS sample. As
expected, star forming galaxies follow a normal distribution in both
the FIR and radio luminosity excess, with a mean $\Delta \log{L}$
value of about zero ($<\Delta \log\mathrm{L_{FIR}}>=0.001$, $<\Delta
\log\mathrm{L_{1.4GHz}}>=0.06$). The root-mean-square scatter is
$0.35$, and $0.32$ for the FIR and radio distributions,
respectively. Assuming the validity of the SFR to radio/FIR luminosity
calibrations, such a (normal) distribution is expected if FIR and
radio emissions arise from star formation processes in the
galaxies. From \f{fig:SFRmainA} \ it is apparent that, as the AGN
contribution (defined via optical spectroscopic emission line
properties) rises in galaxies, the distribution of both the FIR and
radio luminosity excess becomes highly skewed towards higher $\Delta
\log{L}$ values. Although for a fraction of optically selected
  AGN it is possible that the AGN contribution to the radio/FIR may be
  weak ($\Delta \log{L}\sim0$), and they may be overwhelmed by star
  formation (which results in the canonical FIR/radio ratio), the
  significant skewness of the $\Delta \log{L}$ distributions suggests
  an additional source of radio and FIR emission in AGN bearing
  galaxies (at least for $\Delta \log{L}>0$). Even more interesting is
  that galaxies with large luminosity excess in both FIR and radio
  emission ($\Delta \log{L}\gtrsim0.9$) predominantly have FIR/radio
  ratios consistent with the mean $q$ value for star forming galaxies
  (see Fig.~12).  This will be discussed in more detail in
\s{sec:discussion} .

\section{Discussion}
\label{sec:discussion}

\subsection{Comparison with previous results}

Extensive studies of the radio-FIR correlation \citep[e.g.][]{helou85, condon92, yun01, condon02, obric06,mauch07} have led to an average FIR/radio  ratio in the local ($z<0.3$) universe of $q\sim2.3$,
and lower for AGN-bearing galaxies  (see Tab.~2 in
\citealt{sargent10} for a summary). For example, using the IRAS 2~Jy
galaxy sample ($F_\mathrm{60\micron}\geq2$~Jy; 1809 sources with
optical counterparts and well determined redshifts) combined with
 NVSS data, \citet{yun01} have found $<q>=2.34\pm0.01$. A lower average
$q$ value is generally inferred when using
radio selected samples, and reaching fainter in the IR  (see \citealt{sargent10} for a detailed discussion of selection effects). 
Combining NVSS data with the optical Uppsala Galaxy Catalog (UGC) and  
the IRAS FSC and PSC,
\citet{condon02} have found $<q>=2.3$ and rms width $\sigma=0.18$. Furthermore, matching NVSS and 6dFGS survey
data only with the IRAS FSC, \citet{mauch07} inferred a mean $q$ value
of 2.28 with a root-mean-square scatter of 0.22 for their entire sample. For
a subset of radio-loud AGN (that would correspond to
our Seyfert, LINER, absorption, and quasar classes combined) they
found an even lower average value, $<q>=2.0$, and a sigificantly higher
scatter in the FIR/radio ratio ($\sigma=0.5$).

In \f{fig:param} \ we have presented the distribution of $q$ for
various types of our spectroscopically selected NVSS-SDSS-IRAS
(PSC+FSC) galaxies. Our results yield that the dispersion is the
tightest for star forming galaxies ($\sigma=0.13$), and rises by a
factor of 1.5, 2.5, and 2.2 for composite, Seyfert and
absorption/LINER galaxies, respectively. We find that the average
FIR/radio  ratio for all objects in our radio-optical-IR sample
is $2.27\pm0.01$ with a dispersion of $0.2$. This is in very good
agreement with the results from \citet{mauch07}. Furthermore, if we
limit the 60~\micron\ fluxes of our full sample to $\geq2$~Jy we
obtain an average value of 2.34, consistent with that inferred by
\citet{yun01}.

Our results yield a lower FIR/radio ratio ($<q>=2.14\pm0.05$) for
Seyfert galaxies, and a significantly higher root-mean-square-scatter
($\sigma=0.3$), compared to that found for SF galaxies. It is
interesting that the mean $q$ value for our IR-detected LINER and
absorption line galaxies is comparable to that for star forming
galaxies. However, the spread in $q$ for the former is significantly
larger than for the latter (0.28 compared to 0.13, respectively).  The
average FIR/radio ratio for the 21 quasars in our sample is $q=2.04$,
comparable to that inferred for Seyferts and lower than that for star
forming galaxies.  If we combine our AGN-bearing galaxies (quasars,
Seyferts, LINERs, absorption galaxies) into one class in order to
match the AGN sample of \citet{mauch07}, we infer an average $q$ of
$2.16\pm 0.03$ (with a root-mean-square scatter of
$\sigma=0.24$). This is in relatively good agreement with their
results. In the next sections we will discuss the variation of $q$
with radio luminosity and the AGN contribution to the radio-FIR
correlation.

\subsection{ AGN contribution to the radio-FIR correlation }

A low $q$ value is often used to discriminate between star forming
galaxies and AGN. For example \citet{condon02} have classified
radio-loud AGN as those having q $\leq$ 1.8. Assuming that the FIR
emission arises solely from star formation, this criterion selects
galaxies with more than 3 times the radio emission from galaxies on
the FIR-radio correlation. \citet{yun01} have used $q = 1.64$ as a
star formation/AGN separator, identifying galaxies that emit in radio
more than 5 times than predicted by the correlation. It is important
to point out that these discriminating values are tuned to select only
the most radio-loud AGN. Having i) separated our NVSS-SDSS-IRAS sample
into various classes of AGN, and ii) independently estimated SFRs in
their host galaxies, we can now analyze the physical source of FIR and
radio emission in galaxies both following and offset from the
radio-FIR correlation.

Assuming that the additional source of FIR and radio emission
(relative to that expected from star formation) observed in our
composite, Seyfert, absorption and LINER galaxies (see
\f{fig:SFRmainA} \ ) arises from the central supermassive black hole,
the distribution of our luminosity excess, $\Delta \log L$ defined in
eq.~\ref{for:delta}, allows us to constrain the {\em average}
contribution of star formation and AGN activity to the total power
output for a given galaxy population.  Taking that star formation and
AGN activity are the two dominant FIR/radio emission generators, i.e.\
$L_\mathrm{tot} = L_\mathrm{SF} + L_\mathrm{AGN}$, the average
fractional contributions of these two sources ($<f_\mathrm{SF}>$, and
$<f_\mathrm{AGN}>$) to the total power output can then be computed as
$<f_\mathrm{SF}>=10^{-<\Delta \log L>}$, and
$<f_\mathrm{AGN}>=1-<f_\mathrm{SF}>$, where $<\Delta \log L>$ denotes
the average (median) of the $\Delta \log L$ distribution (see
\f{fig:SFRmainA} ).

The median $\Delta \log L$ values, and the fractional star
formation/AGN contributions are summarized in \t{table:fracs} .  As
expected, for star forming galaxies we infer that the average
contribution to FIR and radio emission due to star formation is
$100\%$. We find that composite objects are dominated by star
formation at the $\sim80-90\%$ level.  Further, the FIR emission from
Seyfert galaxies arises predominantly from star formation
($\sim75\%$), while the AGN contribution to radio luminosity in
Seyfert galaxies is about a factor of two higher in the radio than in
the FIR (see \t{table:fracs} ). The latter explains the lower average
$q$ value (compared to the nominal value) for Seyfert galaxies
inferred here, as well as in e.g.\ \citet{obric06},
\citet{mauch07}. Lastly, based on the above calculation IR detected
absorption and LINER galaxies are on average strongly dominated by AGN
activity ($\sim90\%$) in both their FIR and radio emission although
their average FIR/radio ratio is consistent with that expected for
star forming galaxies (see \f{fig:SFRmainA} ).

One of the main results of this work is that, for the large majority
of galaxies with radio and/or IR emission excess, we infer $q$ values
consistent with the average FIR/radio ratio found for star forming
galaxies (see \f{fig:SFRmainA} ). Thus, although a significant AGN
contribution is likely present in these galaxies (adding both to the
FIR and radio emission), they would not be identified with a simple
low-q discriminator value, as is commonly used to select radio-loud
AGN.  Our results indicate that the FIR/radio ratio is not
particularly sensitive to AGN contribution and that the radio-FIR
correlation is a poor discriminant of AGN activity, except for the
most powerful AGN.  This is consistent with observations of several
AGN-bearing galaxies in the high-redshift universe.

Based on an SED analysis, Riechers et al.\ (2009) find that both the
radio and FIR luminosity in the z=3.9 quasar APM08279+5255 are
dominated by the central AGN, but that it has a $q$ value consistent
with the local radio-FIR correlation.  Furthermore, Murphy et al.\ (2009) have
analyzed Spitzer-IRS spectra of a sample of 22 $0.6\lesssim z\lesssim
2.6 $ galaxies, composed of submillimeter galaxies, as well as X-ray
and optically selected AGN in GOODS-N. Making use of their IRS
spectra, they have performed a thorough starburst-AGN decomposition
for each object which allowed them to estimate the fractional AGN
contribution to the total IR luminosity output of each source. They
demonstrate that the 4 galaxies having the largest mid-IR AGN
fractions ($>60\%$) in their sample have $q$ values consistent with
the canonical value. Furthermore, they find that the FIR/radio ratio
shows no trend with the fractional contribution of AGN activity in the
galaxies in the IR, consistent with our results.



\section{Summary and conclusions}
\label{sec:summary}




Based on a new radio-optical-IR catalog we have separated our radio
(NVSS) and IR (IRAS) detected SDSS galaxies ($0.04<z\lesssim0.2$) into
star forming, composite, Seyfert, LINER, absorption line galaxies and
quasars, and we have performed an in-depth analysis of the radio-FIR
correlation for various types of star forming and AGN-bearing
galaxies.  Utilizing our NUV-NIR SED based star formation rates in
combination with FIR and radio luminosity (expected to directly trace
star formation), we have statistically quantified the source of radio
and FIR emission in the galaxies in our sample. We find that Seyfert
galaxies and quasars have FIR/radio ratios lower than the canonical
value for star forming galaxies. This is due to an additional
contribution to their radio continuum emission, which likely arises
from their AGN. We further show that FIR-detected absorption and LINER
galaxies are on average strongly dominated by AGN activity in both
their FIR and radio emission; however their average FIR/radio ratio is
consistent with that expected for star forming galaxies.  In summary,
our results imply that most AGN-containing galaxies in our sample have
FIR/radio flux ratios indistinguishable from those of the star-forming
galaxies. Thus, attempts to separate AGNs from star-forming galaxies
by their FIR/radio flux ratios alone is a poor discriminant of AGN
activity, except for the most powerful radio-loud AGN.


\acknowledgments We are grateful to the anonymous referee for helpful
comments. The authors thank R.~Beck, S.~Charlot, O.~Ilbert,
K.~K.~Knudsen, M.~Sargent, and J.~Walcher for insightful
discussions. IM thanks California Institute of Technology for generous
hospitality. IM and DR acknowledge support from NASA through an award
issued by JPL/Caltech.  VS acknowledges support from the Owens Valley
Radio Observatory, which is supported by the National Science
Foundation through grant AST-0838260. VS \& IM thank Unity through
Knowledge Fund (www.ukf.hr) for collaboration support through the
'Homeland Visit' grant. DR acknowledges support from from NASA through
Hubble Fellowship grant HST-HF-51235.01 awarded by the Space Telescope
Science Institute, which is operated by the Association of
Universities for Research in Astronomy, Inc., for NASA, under contract
NAS 5-26555. AK and ZI ackknowledge NSF grant AST-0507259 to the
University of Washington. The research leading to these results has received funding from the European Union's Seventh Framework programme under grant agreement 229517.

Funding for the SDSS and SDSS-II has been provided by the Alfred P. Sloan Foundation, the Participating Institutions, the National Science Foundation, the U.S. Department of Energy, the National Aeronautics and Space Administration, the Japanese Monbukagakusho, the Max Planck Society, and the Higher Education Funding Council for England. The SDSS Web Site is http://www.sdss.org/.

The SDSS is managed by the Astrophysical Research Consortium for the
Participating Institutions. The Participating Institutions are the
American Museum of Natural History, Astrophysical Institute Potsdam,
University of Basel, University of Cambridge, Case Western Reserve
University, University of Chicago, Drexel University, Fermilab, the
Institute for Advanced Study, the Japan Participation Group, Johns
Hopkins University, the Joint Institute for Nuclear Astrophysics, the
Kavli Institute for Particle Astrophysics and Cosmology, the Korean
Scientist Group, the Chinese Academy of Sciences (LAMOST), Los Alamos
National Laboratory, the Max-Planck-Institute for Astronomy (MPIA),
the Max-Planck-Institute for Astrophysics (MPA), New Mexico State
University, Ohio State University, University of Pittsburgh,
University of Portsmouth, Princeton University, the United States
Naval Observatory, and the University of Washington.

{}

\newpage   
  
\begin{table}[ht]
\caption{Sample summary} 
\centering 
\begin{tabular}{c c c c}
\hline\hline 
 & IRAS (FSC + PSC) & SDSS (MAIN + QUASAR) & IRAS - SDSS \\ 
\hline  \\
 total radio sample & 18313 & 9591 & 524  \\ 
 \hline\hline  \\
 Quasars & -- & 4490 & 21 \\
 \hline\hline  \\
 Absorption & -- & 3072 &  16 \\
 
 Composite & -- & 654 &  203 \\
 
 SF unambiguous  & -- & 621 &  216\\
 
 SF ambiguous  & -- & 9 &  0  \\
 \hline\hline  \\
 AGN unambiguous  & -- & 454 &  43  \\
 
 AGN ambiguous  & -- & 291&  25  \\
 
 Seyfert unambiguous &  -- & 200 & 37\\
 
 LINER unambiguous &  -- & 254 & 6  \\\hline

\hline 

\end{tabular}
\vspace{3mm}
\tablenotetext{0}{The first column denotes the number of radio - IRAS
(Point Source, PS, and Faint Source, FS) catalog with high quality IR
photometry. The second column shows the number of sources in the
radio - SDSS (``main'' and quasar) catalog, and the third column is
the matched radio-SDSS (``main'' and quasar)-IRAS catalog.  The
rows indicate the various galaxy types we separate the objects
into. The unambiguous/ambiguous selection is based on various
 spectroscopic diagnostic tools (see \f{fig:bpts} \ and text for
details). The shown numbers are limited to the $0.04<z<0.3$ redshift range.}


\label{table:class}
\end{table}
   
\newpage 

\begin{table}[ht]
\caption{Fractions of star formation and AGN activity in radio and FIR
  regimes for various types of galaxies} 
\centering 
\begin{tabular}{|c||c|c||c|c|}
\hline\hline
\tablecaption{ Sample summary} 

 &\multicolumn{2}{c|}{RADIO}&\multicolumn{2}{c|}{FIR}\\ 
\cline{2-5}
 &SF&AGN&SF&AGN\\
\hline\hline  
SF&100\%&0\%&100\%&0\%\\ 
Composite&81.3\%&18.7\%&90.7\%&9.3\%\\
Seyfert &56.8\%&43.2\%&76.1\%&23.9\% \\
Abso+LINER&11.3\%&88.7\%&12.8\%&87.2\% \\ [1ex]
\hline

\end{tabular}
\vspace{3mm}

\tablenotetext{0}{Fractional star formation/AGN contribution to the
  radio and FIR for different optically selected galaxy types. For
  each population the
  fractions were estimated in a statistical manner based on the
  distribution of the difference between a galaxy's SED-derived star formation rate
  and radio/FIR luminosity (see 
  \f{fig:SFRmainA} \ and text for details). }

\label{table:fracs}
\end{table}

\newpage

\begin{figure}
\includegraphics[width=\columnwidth]{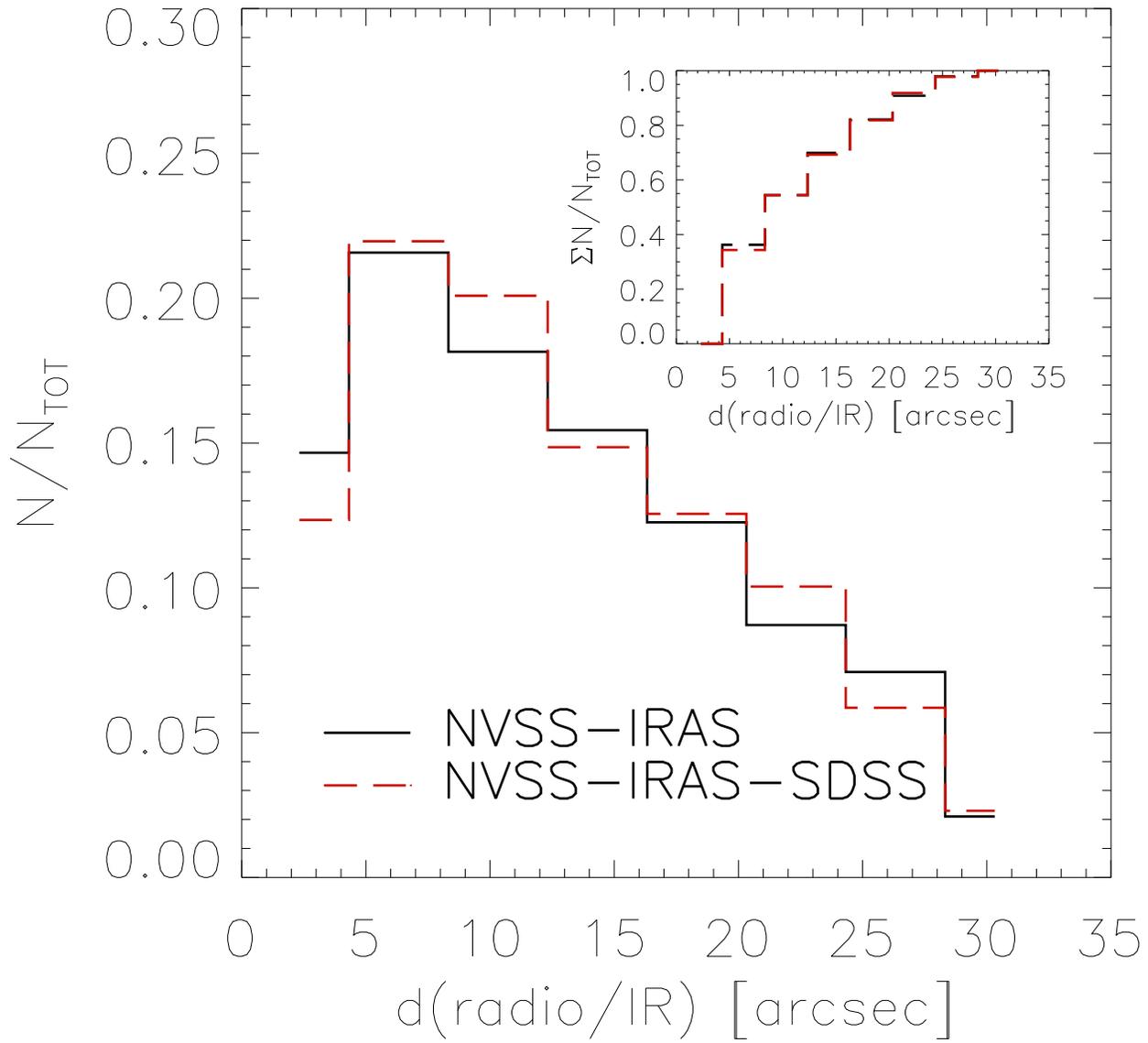}
\caption{Distribution of distances between the radio and FIR detections for the
NVSS-IRAS (full black line) and NVSS-SDSS-IRAS (dashed red line)
samples. The cumulative distribution is shown in the inset.}
	\label{fig:Distance_IRAS_NVSS}
\end{figure}

\newpage

\begin{figure}
\includegraphics[width=\columnwidth]{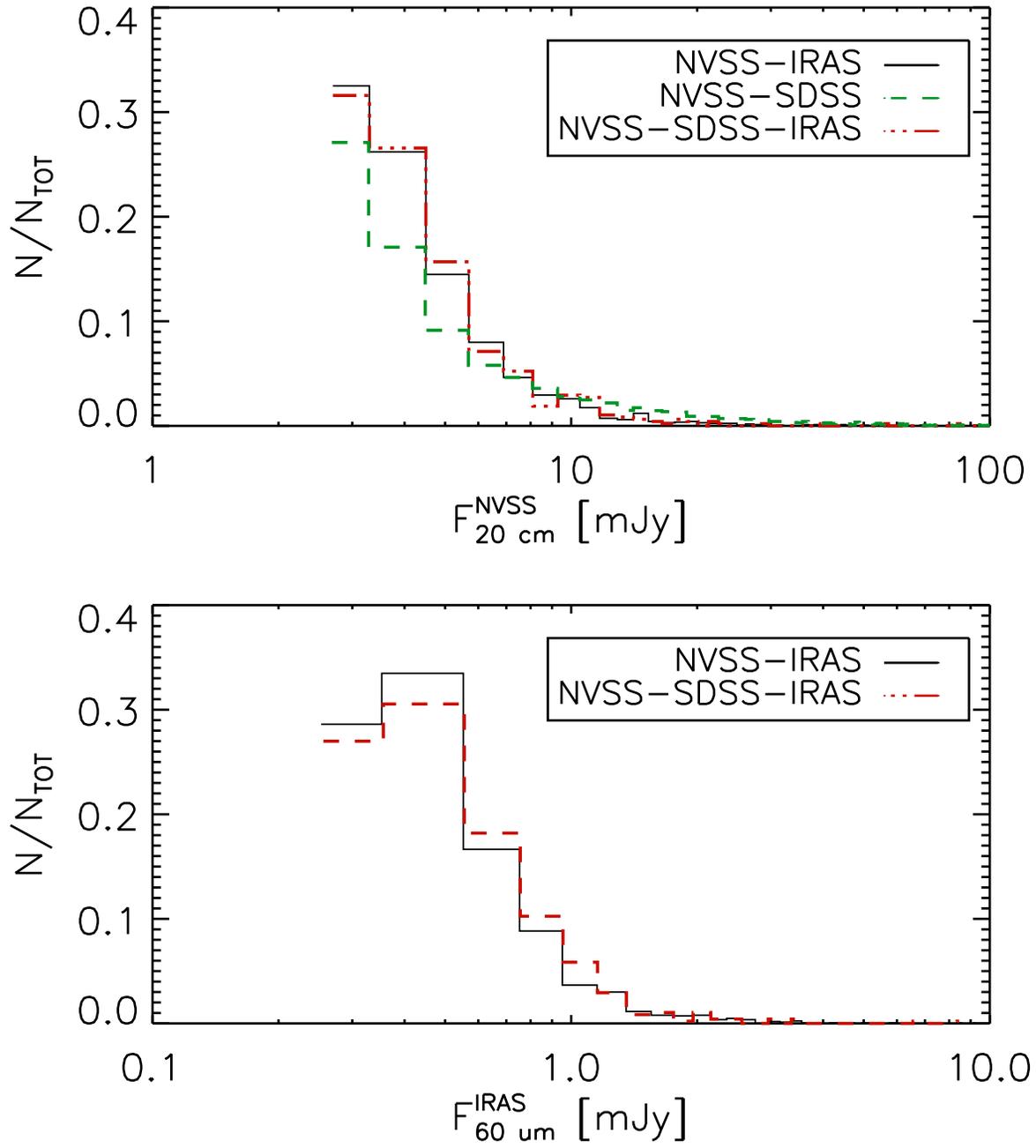}
\caption{The distribution of flux density at 20 cm (top panel) and 60~$\mu$m  (bottom panel) for various radio-selected samples indicated in the top right of the panels.}
	\label{fig:LogFnvss}
\end{figure}

\newpage

\begin{figure}[t!]
\includegraphics[bb = 104 530 486 792, scale=0.7]{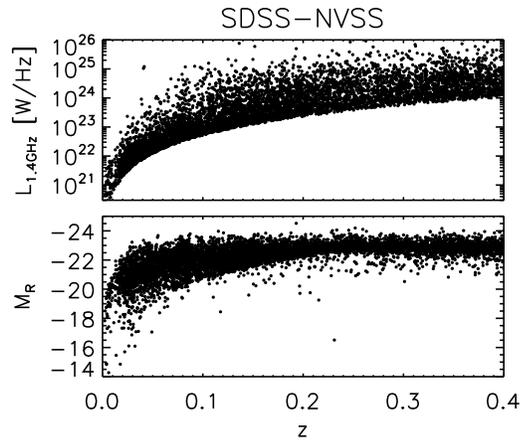}
\caption{ Top panel shows the 1.4GHz luminosity as a function of redshift for the
  NVSS - SDSS galaxies. The bottom panel shows their absolute optical
  r band magnitude (not K-corrected) as a function of redshift.
  \label{fig:limits1}}
\vspace{1.5mm}
\end{figure}

\newpage
\clearpage

\begin{figure}[t!]
\includegraphics[bb = 104 430 486 792, scale=0.7]{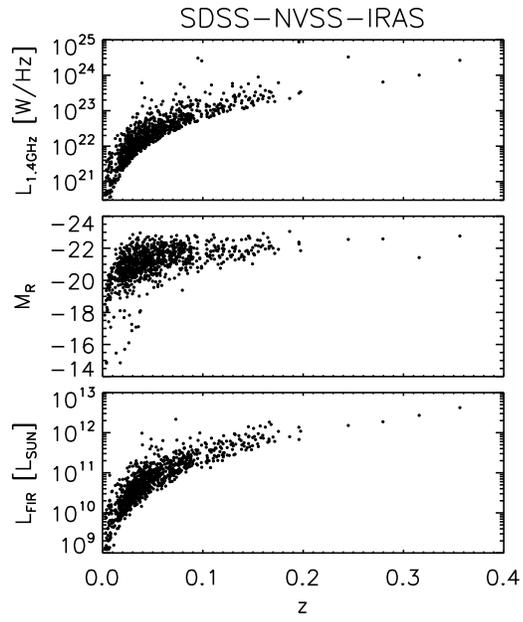}
\caption{ The top two panels are the same as \f{fig:limits1} \ but for
   NVSS - SDSS - IRAS galaxies.  The bottom panel shows the FIR
  luminosity vs.\ redshift. 
  \label{fig:limits2}}
\vspace{1.5mm}
\end{figure}

\newpage

\begin{figure*}
\center{
\includegraphics[bb=54 400 486 830]{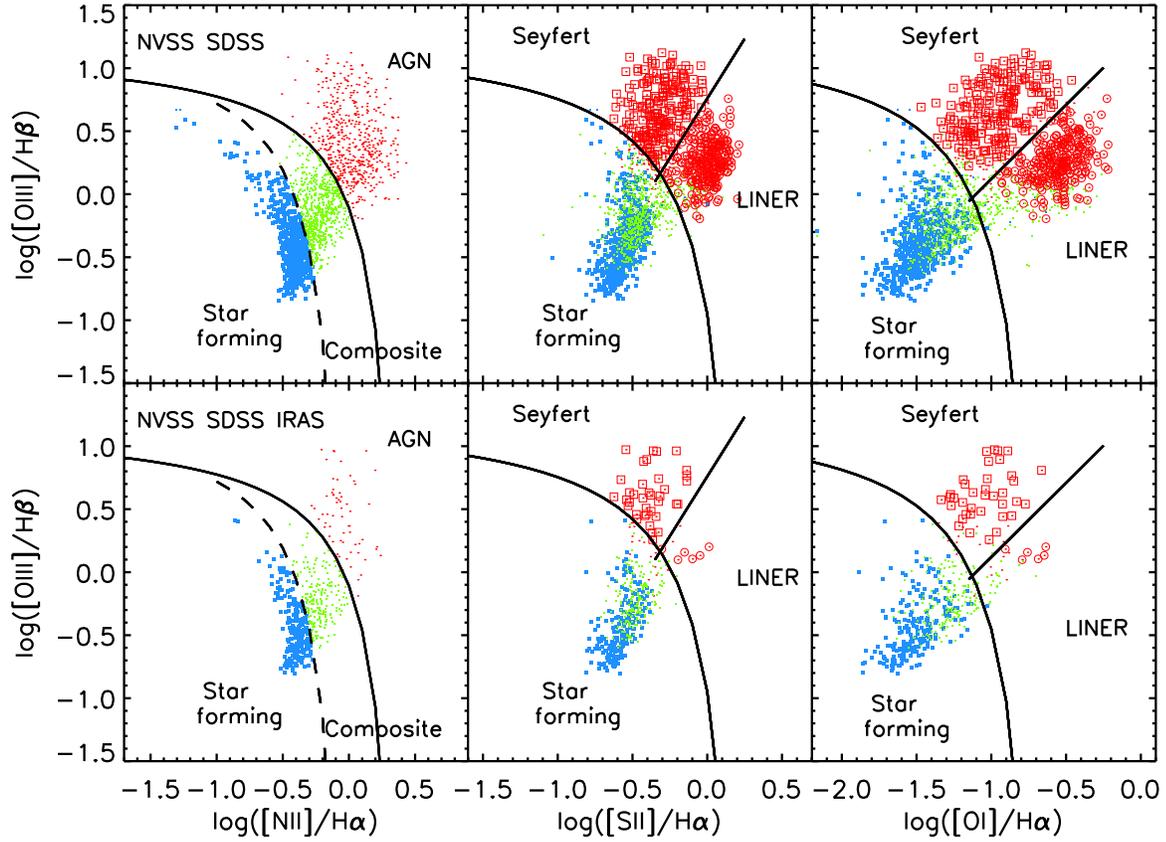}
\caption{ Optical spectroscopic diagnostic diagrams (see
  \citealt{kauffmann03b, kewley01, kewley06}) that separate emission-line
  galaxies into star forming, composite galaxies, and various types of
  AGN (Seyferts and LINERs). The top panel shows the SDSS-NVSS sample,
  and the bottom panel the SDSS-NVSS-IRAS galaxies. Large symbols
  represent unambiguously identified galaxies (see text for details).
  Blue filled squares represent SF galaxies and green dots show
  composites.  Red open squares, and circles represent unambiguous Seyferts and 
  LINER-s, respectively.
  \label{fig:bpts}}
}
\vspace{1.5mm}
\end{figure*}

\newpage

\begin{figure}[t!]
\center{
\includegraphics[width=\columnwidth]{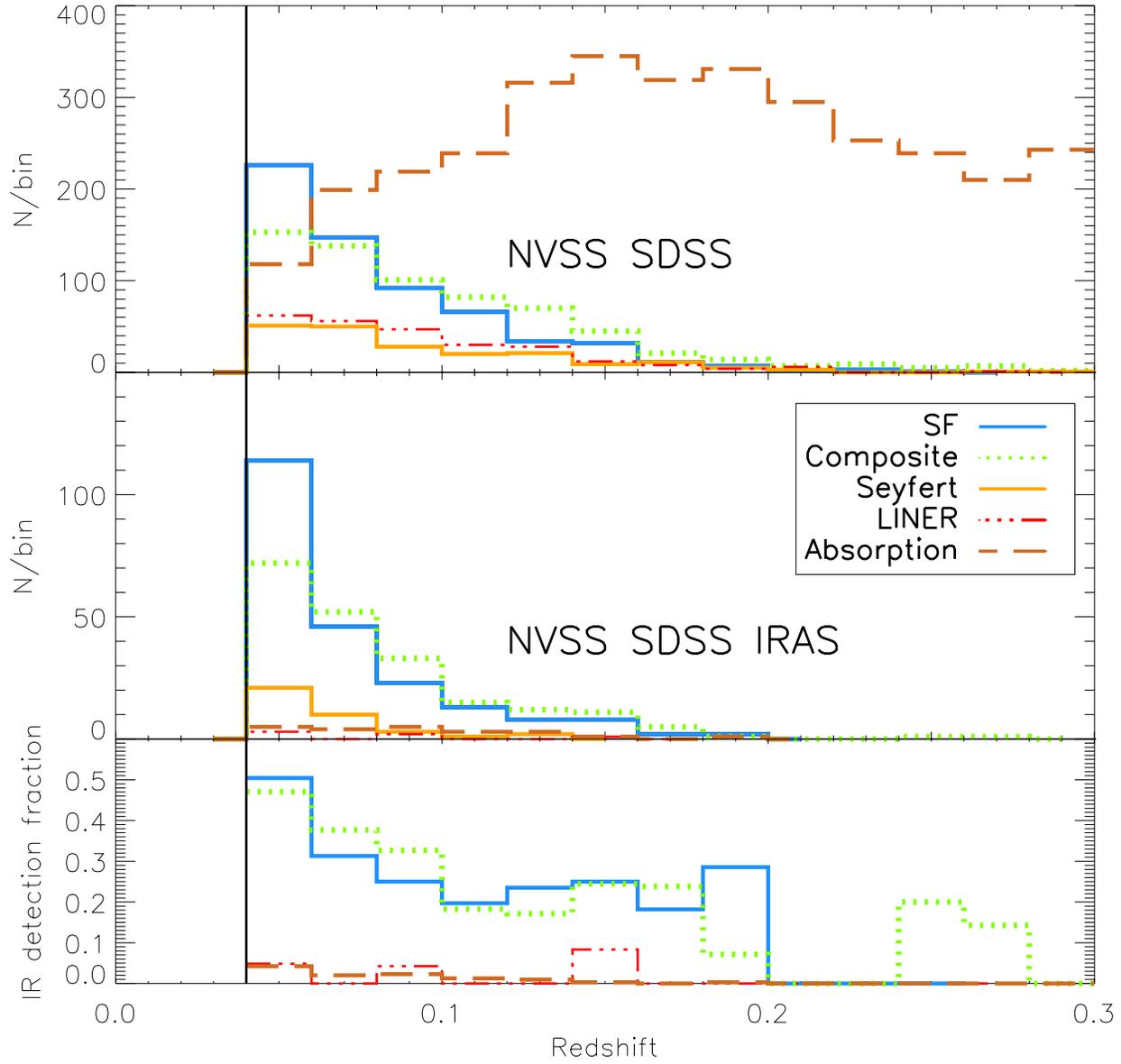}
\caption{ Redshift distribution of galaxies in the NVSS-SDSS
  (top panel) and NVSS-SDSS-IRAS (middle panel) samples. The bottom panel shows the IR detection fraction as a function of redshift. Various galaxy types are indicated in the top right of the middle panel.  
  \label{fig:zhist}}
}
\vspace{1.5mm}
\end{figure}

\newpage
\begin{figure}[t!]
\center{
\includegraphics[bb = 54 370 486 790, width=\columnwidth]{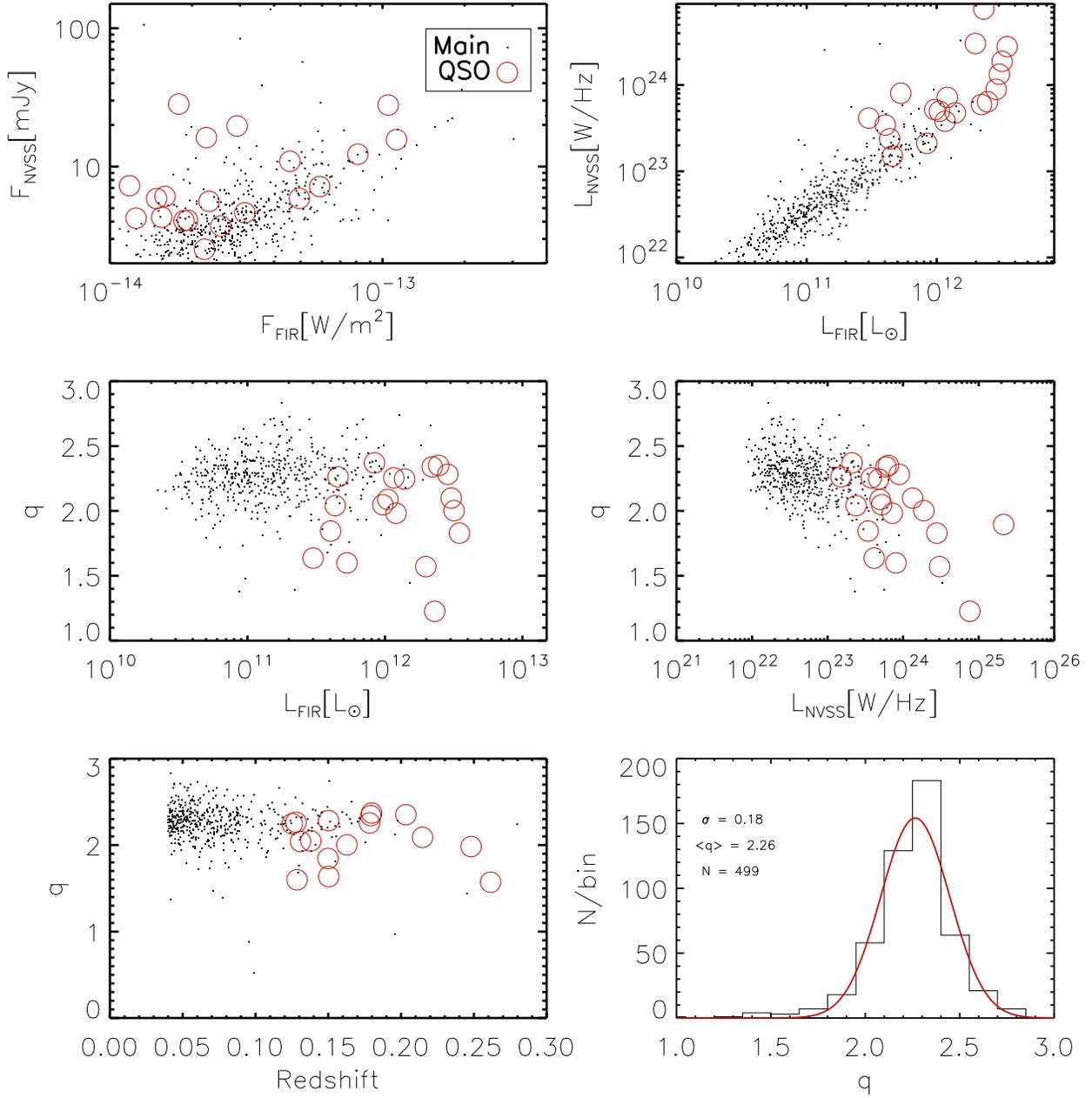}
\caption{ The radio-FIR correlation shown for the NVSS-SDSS-IRAS
  galaxies (dots) and quasars (open circles).  The middle panels show
  the FIR/radio ratio (i.e.\ q parameter defined by eq.~\ref{for:q})
  as a function of FIR (left) and radio (right) luminosities. Note a
  decrease in the q-value with increasing radio luminosity. The bottom
  panels show $q$ as a function of redshift (left) and the
  distribution of $q$ with its best fit Gaussian (right). The number
  of objects in the sample (N), the biweighted mean FIR/radio ratio
  ($<q>$) and the dispersion ($\sigma$) are indicated in the bottom
  right panel.
  \label{fig:IRradio}
}}
\vspace{1.5mm}
\end{figure}

\newpage
\begin{figure}[t!]
\center{
\includegraphics[bb = 54 338 486 790, width=\columnwidth]{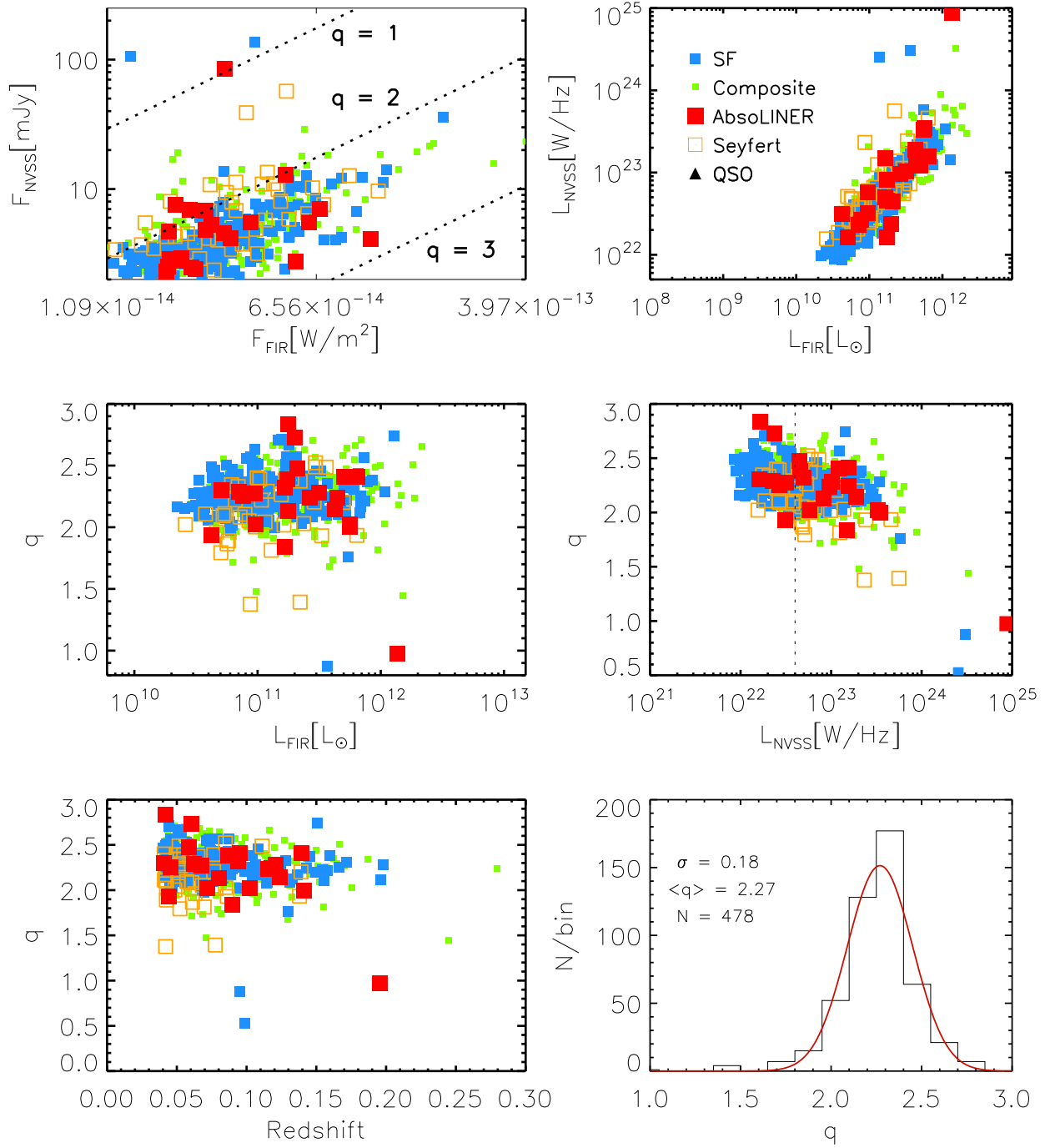}
\caption{ Equivalent to \f{fig:IRradio} , but for different galaxy types. 
  The different symbols are indicated in the top
  right panel, and lines of constant q have been added to the top
  left panel. 
  \label{fig:FIRradio2}}
}
\end{figure}
\newpage

\begin{figure}[t!]
\center{
\includegraphics[bb = 54 380 486 790, width=\columnwidth]{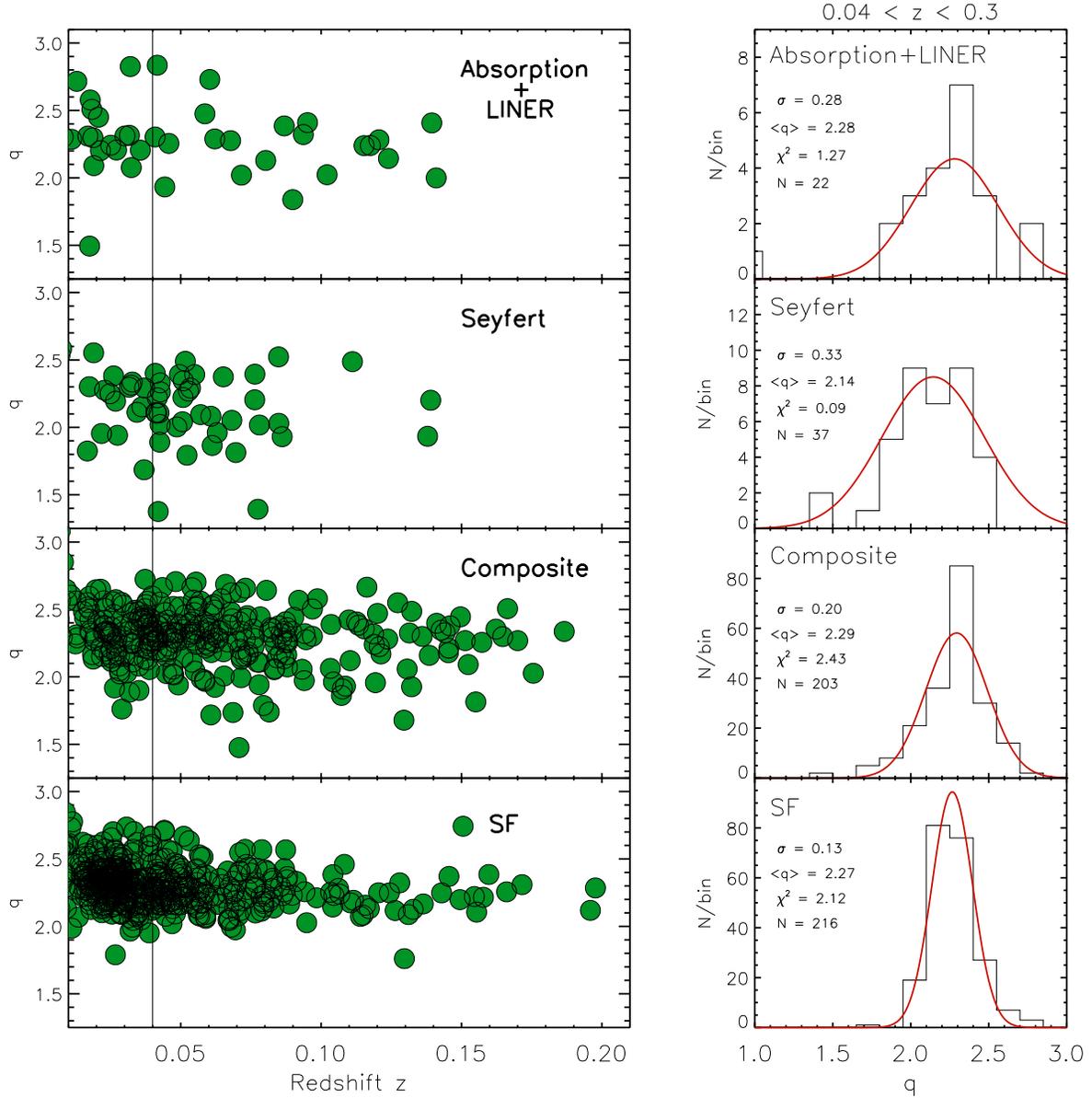}
\caption{ The radio-FIR correlation shown for the NVSS-SDSS-IRAS
  galaxies divided (from top to bottom) into a) absorption line
  galaxies, b) LINERs, c) Seyferts, d) composite, and e) star forming
  galaxies. The left panels show the FIR-radio correlation slope, $q$,
  as a function of redshift. The right panels show the distribution
  of $q$ for each class of galaxies in the redshift range
  $0.04<z<0.3$, free of selection effects due to SDSS fiber-sizes used
  for their optical spectroscopy. 
  \label{fig:param}}
}
\vspace{1.5mm}
\end{figure}
\newpage

\begin{figure}
\center{
\includegraphics[bb = 54 480 486 790, width=\columnwidth]{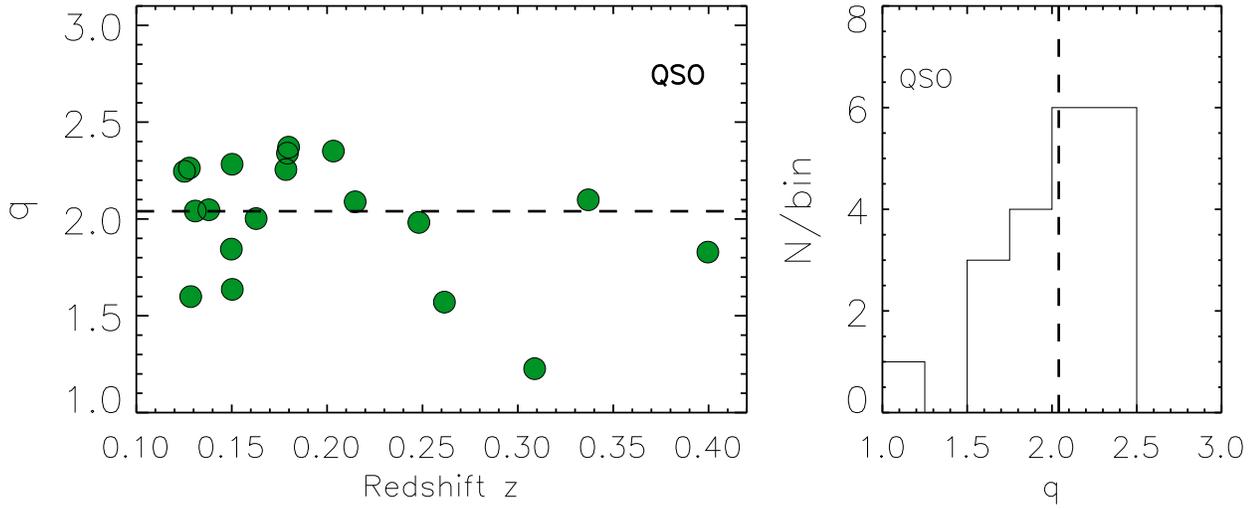}
\caption{ FIR/radio ratio ($q$) as a function of redshift (left panel)
  and the distribution of $q$ for 21 IR (IRAS) detected quasars (right
  panel) in our radio-IR-optical sample. Note that the median $q$ value for qauasars is lower than the average $q$ obtained for star forming galaxies, but comparable to that obtained for Seyfert galaxies (see \f{fig:param} ). The source with $q\sim-1$ is a strong radio source. 
  \label{fig:FIRradioQSO}}
}
\end{figure}
\newpage

\begin{figure*}[th!]
\vspace{15cm}
\includegraphics[bb = 104 360 486 780, scale=0.4]{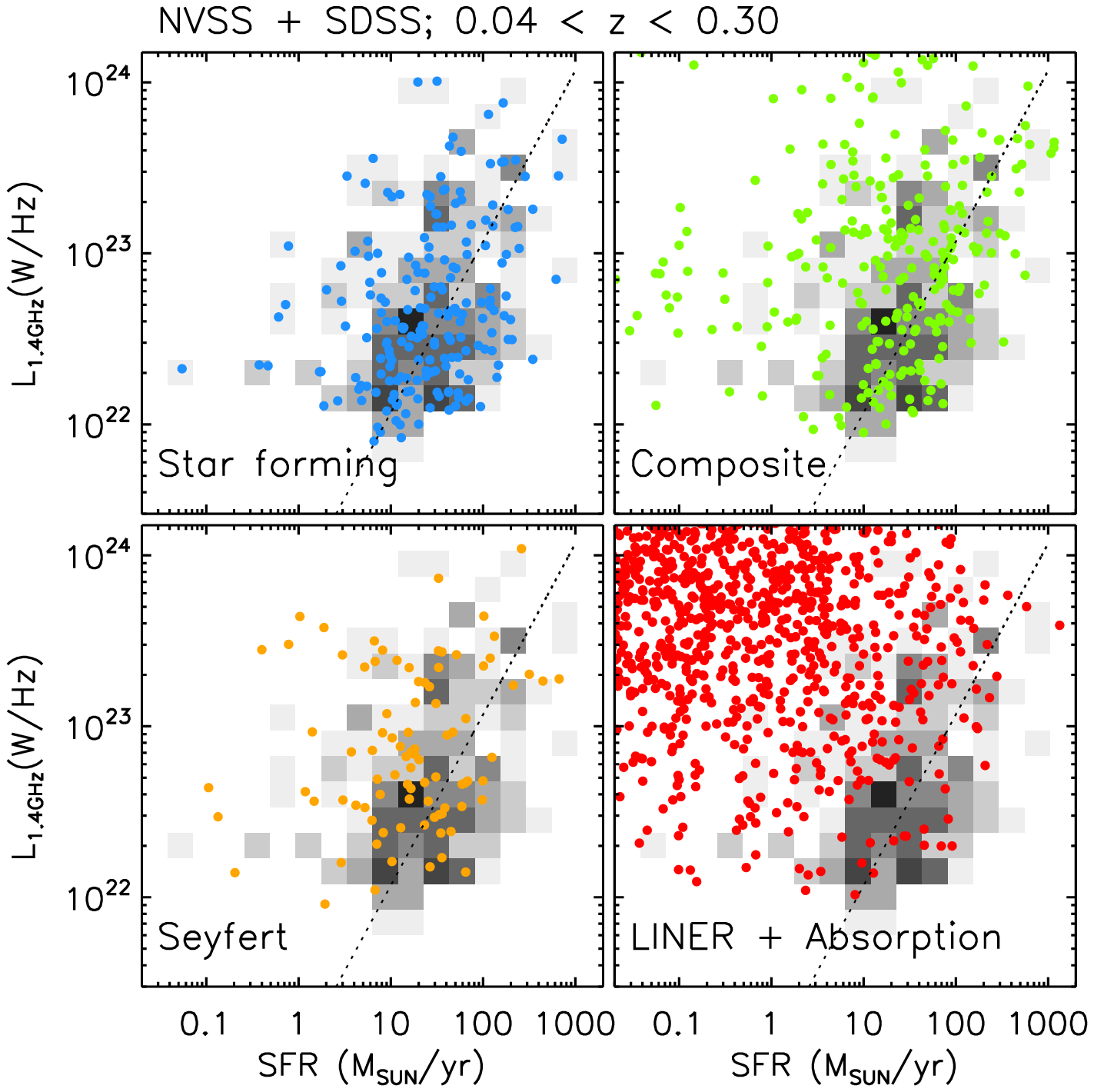}
\includegraphics[bb = 84 360 486 780, scale=0.4]{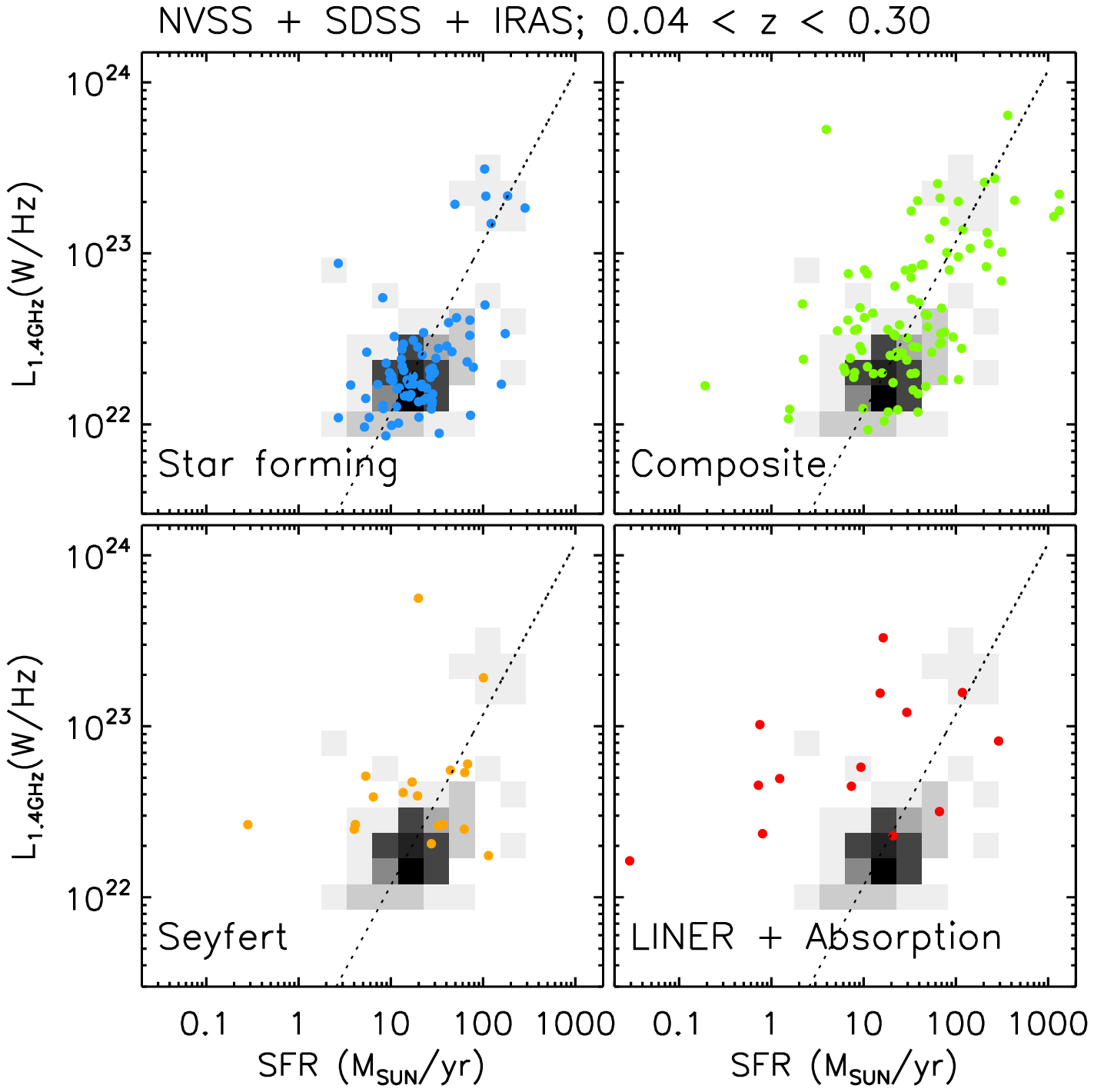}
\includegraphics[bb = 54 360 486 780, scale=0.4]{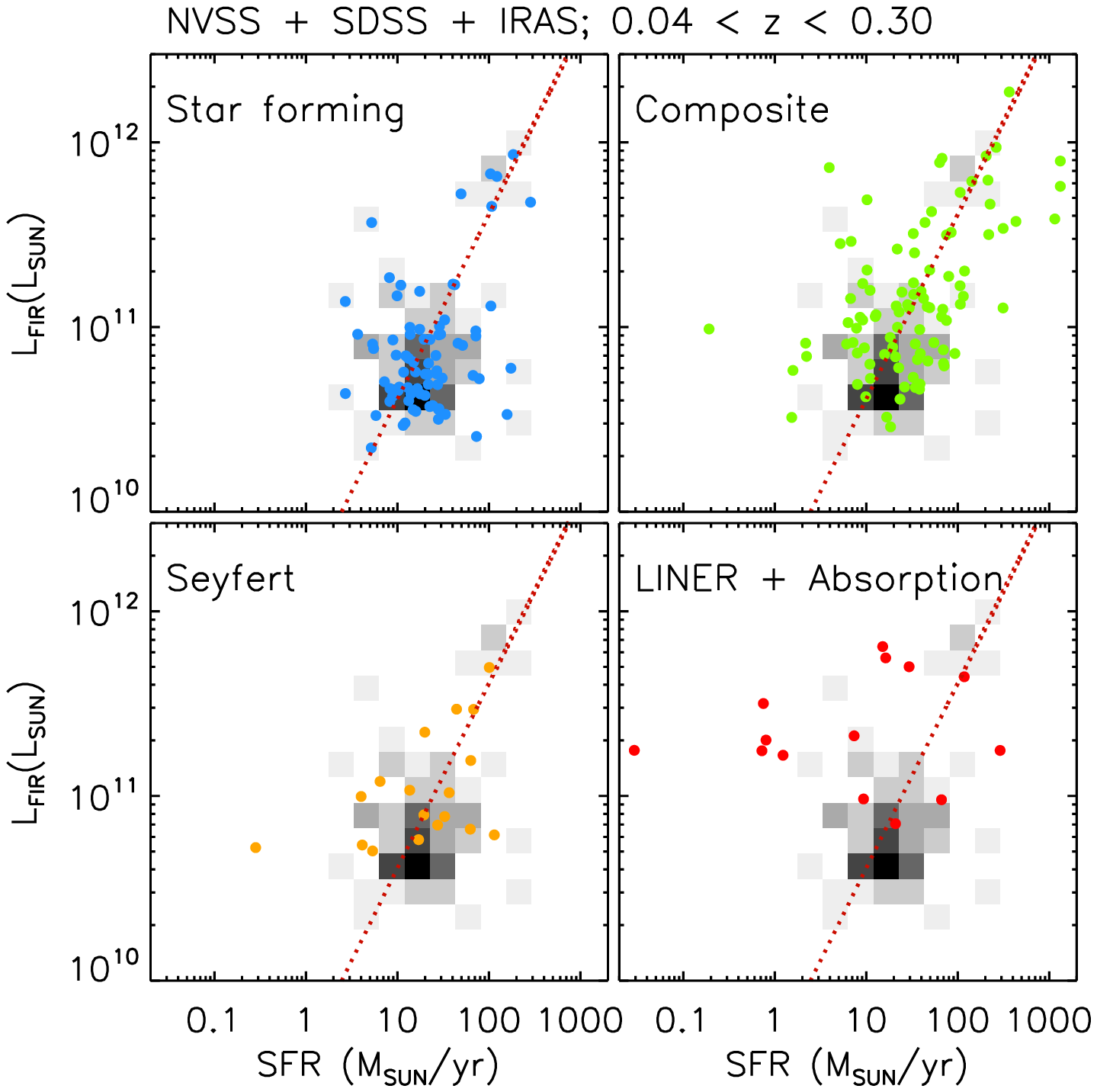}
\caption{ 1.4~GHz radio (left and middle panel) and FIR (right panel)
  luminosity as a function of star formation rate, derived via SED
  fitting to the NUV-NIR SDSS photometry (see text for details). In the
  three large panels different galaxy types (symbols) and samples
  (indicated above and in each panel) are shown. The grey-scale 
  histogram in each plot shows the distribution of star forming
  galaxies for a given sample. Superimposed on the plots (dashed
  lines) are calibrations converting radio and FIR luminosity to star
  formation rate \citep{yun01,kennicutt98}.
  \label{fig:SFRmain}}
\vspace{1.5mm}
\end{figure*}
\newpage

\begin{figure}[h!]
\includegraphics[bb = 54 380 486 790, width=\columnwidth, scale=1.5]{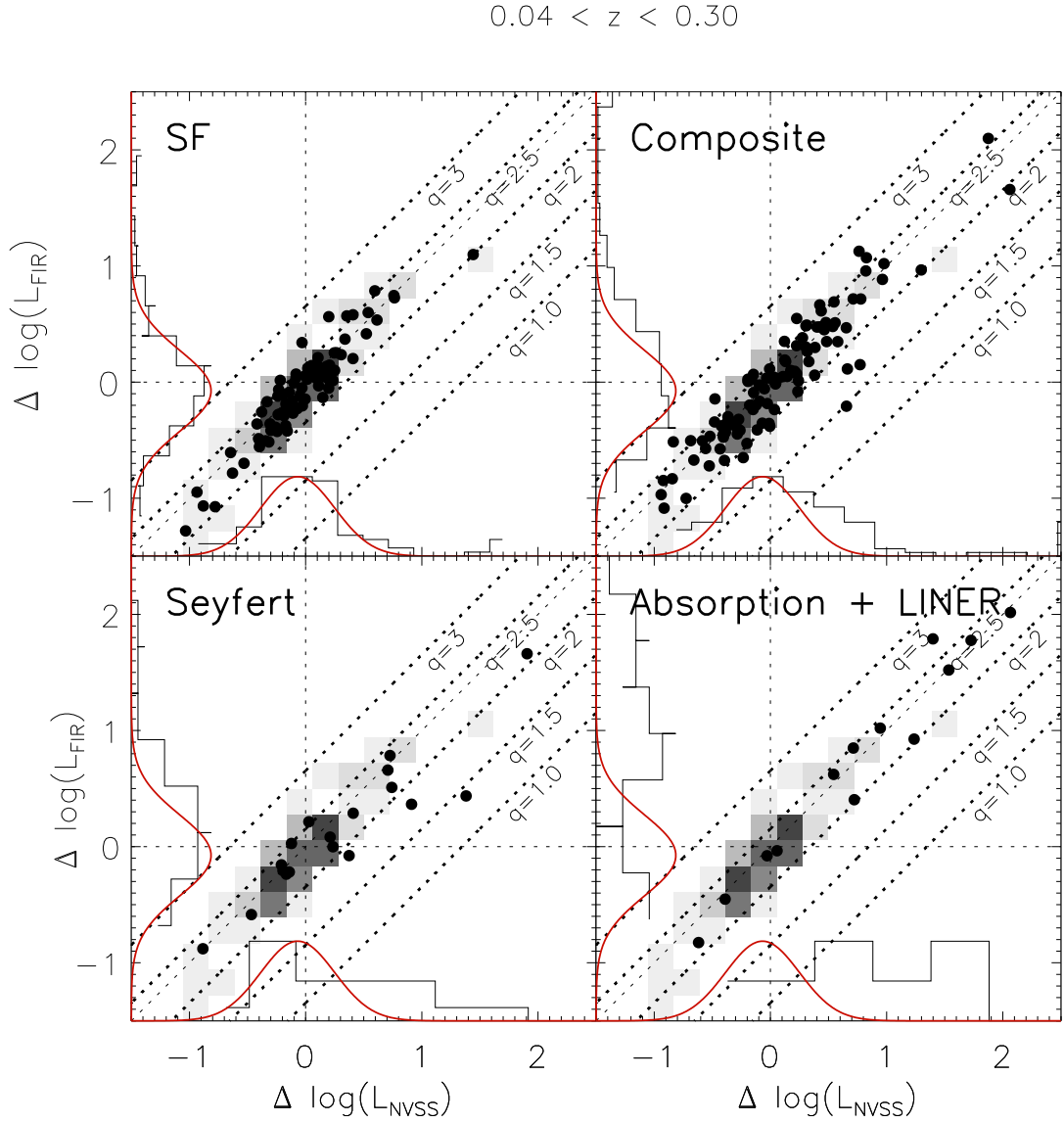}
\caption{ FIR excess ($\Delta \log{\mathrm{L_{FIR}}}$) vs. radio
  excess ($\Delta \log{\mathrm{L_{1.4GHz}}}$) for different types of
  NVSS-SDSS-IRAS galaxies (labeled in each panel) within a redshift
  range of 0.04 to 0.3. The FIR and radio excess has been defined as
  the difference in logarithm of the observed emission and that
  expected from their SFR, as given by the SED modeling (see
  eq.~\ref{for:delta} and \f{fig:SFRmain} \ for details). Histograms
  on the axes show the distributions of the intrinsic
  $\Delta \log{\mathrm{L}}$ values. A Gaussian fit to the distribution
  of {\em star forming} galaxies
  is also shown in each panel (solid line). In each panel filled dots represent the galaxies,
  and the gray-scale shows the distribution of star forming galaxies
  to guide the eye.  Lines of constant q are also shown.
  \label{fig:SFRmainA}}
\vspace{1.5mm}
\end{figure}
\newpage

\end{document}